\documentclass{article}
\usepackage{amssymb}
\usepackage{amsfonts}
\usepackage{amsmath}

\newtheorem{theorem}{Theorem}
\newtheorem{acknowledgement}[theorem]{Acknowledgement}

\begin{document}

\title{Reflectionless $\mathcal{PT}$-symmetric potentials in the
one-dimensional Dirac equation}
\author{Francesco Cannata(1) and Alberto Ventura(1,2) \\
(1) Istituto Nazionale di Fisica Nucleare, Sezione di Bologna, Italy \\ 
 (2) ENEA, Centro Ricerche Ezio Clementel, Bologna, Italy}
\maketitle

\begin{abstract}
We study the one-dimensional Dirac equation with local $\mathcal{PT}$%
-symmetric potentials whose discrete eigenfunctions and continuum asymptotic
eigenfunctions are eigenfunctions of the \ $\mathcal{PT}$ operator, too: on
these conditions the bound-state spectra are real and the potentials are
reflectionless and conserve unitarity in the scattering process. Absence of
reflection makes it meaningful to consider also $\mathcal{PT}$-symmetric
potentials that do not vanish asymptotically.
\end{abstract}

\section{Introduction}

Reflectionless potentials, \textit{i.e.} potentials that are transparent to
incident waves at all energies, have played a special role in quantum
mechanics since the basic paper by Kay and Moses\cite{KM56}, who formulated
the problem of constructing a plane stratified dielectric medium transparent
to electromagnetic radiation in terms of a one-dimensional Schr\H{o}dinger
equation with a potential with preassigned bound-state spectrum that
transmits without reflection continuum wave functions at all incident
energies. From a mathematical point of view, the Kay-Moses method is
equivalent to solving a non-linear Schr\H{o}dinger equation whose potential, 
$V\left( x\right) $, is a quadratic function of a fixed number, $n$, of
unknown bound-state wave functions\cite{NW76}; it can also be considered as
a kind of Hartree-Fock potential with $n$ occupied states for a system of
particles interacting through schematic contact interactions in one space
dimension\cite{NW78}.

More recent approaches to reflectionless potentials in non-relativistic
quantum mechanics make use, among others, of Darboux transformations\cite%
{St95}, supersymmetric hyerarchy derivations from the trivially transparent
constant potential\cite{Ma05} and Casimir invariants of non-compact Lie
groups\cite{KV06}, the latter method giving rise to large families of
reflectionless potentials in implicit form, in addition to explicit
analytical forms derived in previous approaches.

In relativistic quantum mechanics, the Kay-Moses method has been applied to
the one-dimensional Dirac equation with either scalar \cite{NT92}\cite{TNZ93}%
\cite{NT93} or pseudoscalar potentials\cite{NT98}, since the presence of a
vector component may break the transparency of the potential at all energies
(see Ref.\cite{NT92} and Section 3 of the present work), with notable
exceptions, one of which will be discussed in detail in Section 3. The
relativistic extension of the Kay-Moses method is equivalent to the solution
of an auxiliary non-linear Dirac equation.

Reflectionless potentials play an interesting role in non-Hermitian
theories, too, such as quasi-Hermitian quantum mechanics\cite{Ge92}, $%
\mathcal{PT}$-symmetric quantum mechanics\cite{BB98}\cite{Be07}, or
pseudo-Hermitian quantum mecanics\cite{Mo02},\cite{Mo08}. As is known, if a
non-Hermitian potential $V\left( x\right) $ is invariant under the product
of parity $\mathcal{P}$ and time reversal $\mathcal{T}$, so that $\mathcal{PT%
}V\left( x\right) \left( \mathcal{PT}\right) ^{-1}\equiv V^{\ast }\left(
-x\right) =V\left( x\right) $, and the bound-state eigenfunctions of the Schr%
\H{o}dinger Hamiltonian $H=-\frac{1}{2m}\frac{d^{2}}{dx^{2}}+V\left(
x\right) $ are eigenfunctions of $\mathcal{PT}$ (exact $\mathcal{PT}$ \
symmetry), the corresponding eigenvalues are real. As for the continuum of
scattering states, it was proved in Ref.\cite{CDV07} for asymptotically
vanishing potentials in the Schr\H{o}dinger equation that if the asymptotic
wave functions are eigenstates of $\mathcal{PT}$ \ (exact asymptotic $%
\mathcal{PT}$ \ symmetry), the $\mathcal{PT}$-symmetric potential is
reflectionless and unitarity is conserved. In Section 2 of the present work
we extend the proof to the Dirac equation with potentials that admit
non-zero constant limits at $x\rightarrow \pm \infty $.

Scattering from reflectionless potentials with exact asymptotic $\mathcal{PT}
$ \ symmetry can thus be treated by the methods of standard quantum
mechanics, without the need for an equivalent Hermitian formulation, which
is neither exempt from technical difficulties, nor from ambiguity of
interpretation: it has been shown that the equivalent Hermitian description
of scattering from strongly localized non-Hermitian potentials, \ a Dirac
delta function with complex coupling strength in Ref.\cite{Jo07} and a $%
\mathcal{PT}$-symmetric combination of delta functions in Ref. \cite{Jo08},
implies strongly non-local metric operators and, consequently, an apparent
breaking of causality due to incoming waves in the exit channel. This seems
to be the price one has to pay in order to restore unitarity in the
scattering process, \ although a new formulation of the problem\cite{Zn08a}%
\cite{Zn08b}, based on the discretization of the Schr\H{o}dinger equation on
an infinite one-dimensional lattice, has provided examples where the metric
operator in the Hermitian equivalent formulation can be chosen as a diagonal
matrix, called a quasi-local operator, which prevents the appearance of
incoming waves in the exit channel, at the cost of a change of scale of the
probability density on the left and the right of the scattering centre.

In the present state of formulation of quasi-Hermitian theories, however, we
share the opinion expressed in Ref.\cite{Jo08}, that it makes sense to treat
a non-Hermitian scattering potential as an effective one, accepting that it
may well involve the loss of unitarity when attention is restricted to the
system itself and not its environment, with which it can exchange
probability flux (see also Refs. \cite{CJT98}\cite{ACDJ99}\cite{SAC06}). On
the other hand, reflectionless potentials are a special class of $\mathcal{PT%
}$-symmetric potentials that conserve unitarity even in the standard
formulation of quantum mechanics; therefore, we believe that they may
deserve a study of their own, not only in the standard framework, \textit{%
i.e.} with a trivial metric operator, adopted here, since it does not give
rise to unphysical aspects for an isolated system, but possibly also as a
test of alternative approaches.

The main scope of the present work is to investigate the behaviour of
reflectionless potentials in relativistic quantum mechanics, under different
conditions of Lorentz covariance, \textit{\ i.e. }when they appear as
vector, scalar or pseudoscalar components in the one-dimensional Dirac
equation. Section 2 describes the general formalism, examples of
scalar-plus-vector potentials are worked out in Section 3, pseudoscalar
potentials in Section 4 and scalar potentials in Section 5. Finally, Section
6 is dedicated to conclusions and perspectives.

\section{General formalism}

The time-independent Dirac equation in (1+1) dimensions with vector , scalar
and pseudoscalar potentials, $V$, $S$ and $P$, respectively, reads, in units 
$\hbar =c=1$%
\begin{equation}
\left[ \alpha _{x}p_{x}+\beta \left( m+S(x\right) )+i\alpha _{x}\beta
P\left( x\right) +V(x)\right] \Psi \left( x\right) =E\Psi \left( x\right) \;.
\label{gen_Dir_eq}
\end{equation}

Here, $\Psi \left( x\right) $ is a two-dimensional spinor and $p_{x}\equiv -i%
\frac{d}{dx}$. $\alpha _{x}$ and $\beta $ are two anti-commuting Hermitian
traceless matrices with the property $\alpha _{x}^{2}=\beta ^{2}=\left( 
\begin{array}{cc}
1 & 0 \\ 
0 & 1%
\end{array}%
\right) \equiv 1_{2}$, which can be identified with two Pauli matrices.

We assume for generality's sake that $S$, $P$ and $V$ can have non-zero
limits at $x=\pm \infty $: $\lim_{x\rightarrow \pm \infty }S(x)=S_{\pm }$,
and analogous notations for $P$ and $V$. In the Dirac representation\cite%
{MS87}, $\alpha _{x}=\sigma _{x}=\left( 
\begin{array}{cc}
0 & 1 \\ 
1 & 0%
\end{array}%
\right) $ and $\beta =\sigma _{z}=\left( 
\begin{array}{cc}
1 & 0 \\ 
0 & -1%
\end{array}%
\right) $, and the asymptotic Dirac equation reads%
\begin{equation}
\left( 
\begin{array}{cc}
m+S_{\pm }+V_{\pm }-E & -i\left( \frac{d}{dx}+P_{\pm }\right) \\ 
i\left( -\frac{d}{dx}+P_{\pm }\right) & -m-S_{\pm }+V_{\pm }-E%
\end{array}%
\right) \left( 
\begin{array}{c}
\psi _{1}\left( x\right) \\ 
\psi _{2}\left( x\right)%
\end{array}%
\right) =0\;.  \label{asympt_Dir_eq}
\end{equation}

Let us search for a solution of the following form%
\begin{equation}
\begin{array}{c}
\psi _{1}\left( x\right) =A_{\pm }e^{ik_{\pm }x}+B_{\pm }e^{-ik_{\pm }x} \\ 
\psi _{2}\left( x\right) =A_{\pm }C_{\pm }e^{ik_{\pm }x}+B_{\pm }D_{\pm
}e^{-ik_{\pm }x}%
\end{array}%
\;,  \label{psi_trial}
\end{equation}%
where $A_{\pm }$, $B_{\pm }$ $C_{\pm }$ and $D_{\pm }$ are complex numbers.

Direct substitution of formulae (\ref{psi_trial}) in the asymptotic Dirac
equation (\ref{asympt_Dir_eq}) yields%
\begin{equation*}
C_{\pm }=\frac{k_{\pm }+iP_{\pm }}{m+S_{\pm }+E-V_{\pm }}\;,\;D_{\pm }=\frac{%
-k_{\pm }+iP_{\pm }}{m+S_{\pm }+E-V_{\pm }}
\end{equation*}%
and the asymptotic momenta satisfy the relation 
\begin{equation}
k_{\pm }^{2}=\left( E-V_{\pm }\right) ^{2}-\left( m+S_{\pm }\right)
^{2}-P_{\pm }^{2}\;,  \label{asympt_k}
\end{equation}%
while $A_{\pm }$ and $B_{\pm }$ remain to be fixed on boundary conditions.

Like in our previous works\cite{CV08a},\cite{CV08b}, which use the same
representation of Dirac matrices, the parity operator, $\mathcal{P}$, and
the time reversal operator, $\mathcal{T}$, are%
\begin{equation*}
\begin{array}{c}
\mathcal{P}=P_{0}\beta =P_{0}\sigma _{z}\;, \\ 
\mathcal{T}=\mathcal{K}\beta =\mathcal{K}\sigma _{z}\;,%
\end{array}%
\end{equation*}%
where $P_{0}$ changes $x$ into $-x$ and $\mathcal{K}$ performs complex
conjugation. Their product%
\begin{equation}
\mathcal{PT}=P_{0}\mathcal{K}\sigma _{z}^{2}=P_{0}\mathcal{K}  \label{PT}
\end{equation}%
is thus the same as in non-relativistic quantum mechanics\cite{CDV07}.

If the Dirac Hamiltonian on the left-hand side of Eq. (\ref{gen_Dir_eq}) is $%
\mathcal{PT}$-symmetric, $S_{+}=S_{-}^{\ast }$ ,$\ $\ $V_{+}=V_{-}^{\ast }$
and $P_{+}=-P_{-}^{\ast }$ . $\mathcal{PT}$ symmetry of the potentials thus
implies $k_{\pm }^{2\ast }=k_{\mp }^{2}$, which means that either $%
k_{-}^{\ast }=k_{+}$, or $k_{-}^{\ast }=-k_{+}$. Using formulae (\ref%
{psi_trial}), it is easy to show that only with the choice $k_{-}^{\ast
}=k_{+}$ the ratios of transmitted waves over incident waves remain finite
even if the amplitudes of asymptotic wave functions may diverge at $x=\pm
\infty $ . This argument does not hold for the reflection coefficient:
therefore, only when reflection is identically zero it makes sense to treat $%
\mathcal{PT}$-symmetric potentials that do not vanish asymptotically. In
turn, $k_{\pm }^{\ast }=k_{\mp }$ implies $C_{\pm }^{\ast }=C_{\mp }$ and $%
D_{\pm }^{\ast }=D_{\mp }$. When all the potentials vanish asymptotically,
the well-known expressions for free particles are recovered: $%
k^{2}=E^{2}-m^{2}$, $C_{\pm }=\frac{k}{E+m}$, $D_{\pm }=-\frac{k}{E+m}$. \ 

In general, $A_{\pm }$ and $B_{\pm }$ are linear combinations of the
coefficients of asymptotic expansions of two linearly independent solutions
to Eq. (\ref{gen_Dir_eq}), $\Psi ^{\left( 1\right) }\left( x\right) $ and $%
\Psi ^{\left( 2\right) }\left( x\right) $%
\begin{equation*}
\lim_{x\rightarrow \pm \infty }\Psi ^{\left( i\right) }\left( x\right)
=a_{i\pm }\left( k_{\pm }\right) \left( 
\begin{array}{c}
1 \\ 
C_{\pm }%
\end{array}%
\right) e^{ik_{\pm }x}+b_{i\pm }\left( k_{\pm }\right) \left( 
\begin{array}{c}
1 \\ 
D_{\pm }%
\end{array}%
\right) e^{-ik_{\pm }x}\;\;\left( i=1,2\right)
\end{equation*}%
in the general asymptotic solution%
\begin{equation*}
\lim_{x\rightarrow \pm \infty }\Psi \left( x\right) =\alpha
\lim_{x\rightarrow \pm \infty }\Psi ^{\left( 1\right) }\left( x\right)
+\beta \lim_{x\rightarrow \pm \infty }\Psi ^{\left( 2\right) }\left(
x\right) \;,
\end{equation*}%
or%
\begin{equation*}
\begin{array}{c}
A_{\pm }=\alpha a_{1\pm }\left( k_{\pm }\right) +\beta a_{2\pm }\left(
k_{\pm }\right) \;, \\ 
B_{\pm }=\alpha b_{1\pm }\left( k_{\pm }\right) +\beta b_{2\pm }\left(
k_{\pm }\right) \;.%
\end{array}%
\end{equation*}

$\alpha $ and $\beta $, in turn, can be fixed by boundary conditions. If $%
\Psi \left( x\right) $ is a progressive wave, travelling from left to right,
we must have, apart from a global normalization constant, not relevant in
this context,%
\begin{equation*}
\left\{ 
\begin{array}{c}
\lim_{x\rightarrow -\infty }\Psi \left( x\right) =\left( 
\begin{array}{c}
1 \\ 
C_{-}%
\end{array}%
\right) e^{ik_{-}x}+R_{L\rightarrow R}\left( 
\begin{array}{c}
1 \\ 
D_{-}%
\end{array}%
\right) e^{-ik_{-}x}\;, \\ 
\lim_{x\rightarrow +\infty }\Psi \left( x\right) =T_{L\rightarrow R}\left( 
\begin{array}{c}
1 \\ 
C_{+}%
\end{array}%
\right) e^{ik_{+}x}\;,%
\end{array}%
\right.
\end{equation*}%
where the transmission and reflection coefficients, $T_{L\rightarrow R}$ and 
$R_{L\rightarrow R}$, have been introduced.

Therefore%
\begin{equation*}
\left\{ 
\begin{array}{c}
A_{-}=\alpha a_{1-}+\beta a_{2-}=1\;, \\ 
B_{-}=\alpha b_{1-}+\beta b_{2-}=R_{L\rightarrow R}\;, \\ 
B_{+}=\alpha b_{1+}+\beta b_{2+}=0\;, \\ 
A_{+}=\alpha a_{1+}+\beta a_{2+}=\;T_{L\rightarrow R}\;\ ,%
\end{array}%
\right.
\end{equation*}%
whence%
\begin{equation}
\left\{ 
\begin{array}{c}
T_{L\rightarrow R}=\frac{b_{1+}a_{2+}-a_{1+}b_{2+}}{b_{1+}a_{2-}-a_{1-}b_{2+}%
}\;, \\ 
R_{L\rightarrow R}=\frac{b_{1+}b_{2-}-b_{1-}b_{2+}}{b_{1+}a_{2-}-a_{1-}b_{2+}%
}\;.%
\end{array}%
\right.  \label{TR_LR}
\end{equation}

$\;.$

In the same way, if $\Psi \left( x\right) $ is a regressive wave, travelling
from right to left%
\begin{equation*}
\left\{ 
\begin{array}{c}
\lim_{x\rightarrow -\infty }\Psi \left( x\right) =T_{R\rightarrow L}\left( 
\begin{array}{c}
1 \\ 
D_{-}%
\end{array}%
\right) e^{-ik_{-}x}\;, \\ 
\lim_{x\rightarrow +\infty }\Psi \left( x\right) =\left( 
\begin{array}{c}
1 \\ 
D_{+}%
\end{array}%
\right) e^{-ik_{+}x}+R_{R\rightarrow L}\left( 
\begin{array}{c}
1 \\ 
C_{+}%
\end{array}%
\right) e^{ik_{+}x}\;.%
\end{array}%
\right.
\end{equation*}

Thus%
\begin{equation*}
\left\{ 
\begin{array}{c}
A_{-}=\alpha a_{1-}+\beta a_{2-}=0\;, \\ 
B_{-}=\alpha b_{1-}+\beta b_{2-}=T_{R\rightarrow L}\;, \\ 
A_{+}=\alpha a_{1+}+\beta a_{2+}=R_{R\rightarrow L}\;, \\ 
B_{+}=\alpha b_{1+}+\beta b_{2+}=1\;.%
\end{array}%
\right.
\end{equation*}%
whence%
\begin{equation}
\left\{ 
\begin{array}{c}
T_{R\rightarrow L}=\frac{a_{2-}b_{1-}-a_{1-}b_{2-}}{%
b_{1+}a_{2--}-a_{1-}b_{2+}}\;, \\ 
R_{R\rightarrow L}=\frac{a_{1+}a_{2-}-a_{1-}a_{2+}}{b_{1+}a_{2-}-a_{1-}b_{2+}%
}\;.%
\end{array}%
\right.  \label{TR_RL}
\end{equation}

Not surprisingly, the transmission and reflection coefficients (\ref{TR_LR})
and (\ref{TR_RL}) are the same as in the non-relativistic case\cite{CDV07}.

The Wronskian of two solutions of the Dirac equation, $\Psi ^{\left(
1\right) }\left( x\right) \equiv \left( 
\begin{array}{c}
\psi _{1}^{\left( 1\right) }\left( x\right) \\ 
\psi _{2}^{\left( 1\right) }\left( x\right)%
\end{array}%
\right) $ and $\Psi ^{\left( 2\right) }\left( x\right) \equiv \left( 
\begin{array}{c}
\psi _{1}^{\left( 2\right) }\left( x\right) \\ 
\psi _{2}^{\left( 2\right) }\left( x\right)%
\end{array}%
\right) $ is defined as%
\begin{equation}
W\left( x\right) \equiv \left\vert 
\begin{array}{cc}
\psi _{1}^{\left( 1\right) }\left( x\right) & \psi _{1}^{\left( 2\right)
}\left( x\right) \\ 
\psi _{2}^{\left( 1\right) }\left( x\right) & \psi _{2}^{\left( 1\right)
}\left( x\right)%
\end{array}%
\right\vert =\psi _{1}^{\left( 1\right) }\left( x\right) \psi _{2}^{\left(
1\right) }\left( x\right) -\psi _{1}^{\left( 2\right) }\left( x\right) \psi
_{2}^{\left( 1\right) }\left( x\right) \;.  \label{Wronsk}
\end{equation}

It is easy to check that $\frac{dW\left( x\right) }{dx}=0$, \textit{i.e. }$%
W\left( x\right) =const.$, by expressing the derivatives of the spinor
components as linear combinations of the components themselves, as dictated
by Eq. (\ref{gen_Dir_eq}). If the two solutions are linearly independent, $%
W\neq 0$, of course.

Using definition (\ref{Wronsk}) and asymptotic wave functions, $\Psi
^{\left( i\right) }{}_{\pm }\equiv \lim_{x\rightarrow \pm \infty }\Psi
^{\left( i\right) }\left( x\right) $, one easily obtains 
\begin{equation}
\lim_{x\rightarrow \pm \infty }W\left( x\right) \equiv W_{\pm }=\left(
a_{1\pm }b_{2\pm }-a_{2\pm }b_{1\pm }\right) \left( D_{\pm }-C_{\pm }\right)
\;.  \label{W_pm}
\end{equation}

Remembering expressions (\ref{TR_LR}-\ref{TR_RL}) of the transmission
coefficients, formula (\ref{W_pm}) yields%
\begin{equation*}
\begin{array}{c}
W_{-}=\left( a_{1-}b_{2+}-a_{2-}b_{1+}\right) T_{R\rightarrow
L}(D_{-}-C_{-})\;, \\ 
W_{+}=\left( a_{1-}b_{2+}-a_{2-}b_{1+}\right) T_{L\rightarrow
R}(D_{+}-C_{+})\;.%
\end{array}%
\end{equation*}%
and $W_{-}=W_{+}$ is equivalent to%
\begin{equation*}
T_{R\rightarrow L}(D_{-}-C_{-})=T_{L\rightarrow R}(D_{+}-C_{+})\;,
\end{equation*}%
or%
\begin{equation}
\frac{T_{L\rightarrow R}}{T_{R\rightarrow L}}=\frac{D_{-}-C_{-}}{D_{+}-C_{+}}%
=\frac{D_{-}-C_{-}}{D_{-}^{\ast }-C_{-}^{\ast }}=e^{i\nu }\;,  \label{Ratio}
\end{equation}

Here, $\nu =2\arg \left( D_{-}-C_{-}\right) $ is a real phase. When, in
particular, all potentials vanish asymptotically, $C_{-}=-D_{-}$ are real
numbers and the two transmission coefficients are equal.

The phase difference, $\nu ,$ of the two transmission coefficients is
different from zero when the imaginary components of the $\mathcal{PT}$%
-symmetric potentials do not vanish asymptotically and is present in
non-relativistic quantum mechanics, too, as recently shown in Ref.\cite{LM09}
for a $\mathcal{PT}$-symmetric version of the hyperbolic Rosen-Morse
potential.

It is worthwhile to point out that the formalism just developed refers to
local potentials. For non-local potentials it has been shown that the ratio
of the two transmission coefficients is not 1, but a complex number of unit
modulus, even if the imaginary potentials vanish asymptotically, both in
non-relativistic\cite{CV06} and relativistic wave equations\cite{CV08b}. In
this case the two reflection coefficients have the same phase, but different
modulus and unitarity is broken.

Let us now apply the $\mathcal{PT}$ operator (\ref{PT}) to the general
asymptotic wave functions%
\begin{equation*}
\lim_{x\rightarrow \pm \infty }\Psi \left( x\right) \equiv \Psi _{\pm
}\left( x\right) =A_{\pm }\left( 
\begin{array}{c}
1 \\ 
C_{\pm }%
\end{array}%
\right) e^{ik_{\pm }x}+B_{\pm }\left( 
\begin{array}{c}
1 \\ 
D_{\pm }%
\end{array}%
\right) e^{-ik_{\pm }x}\;\;,
\end{equation*}%
or, more conveniently, to the following interpolating function, which
coincides with the asymptotic wave functions at large $\left\vert
x\right\vert $%
\begin{equation*}
\Psi _{int.}\left( x\right) =\frac{1}{2}(1+sgn(x))\Psi _{+}\left( x\right) +%
\frac{1}{2}(1-sgn(x))\Psi _{-}\left( x\right) \;.
\end{equation*}

By definition (\ref{PT}) one gets%
\begin{equation*}
\mathcal{PT}\Psi _{int.}\left( x\right) =\Psi _{int.}^{\ast }\left(
-x\right) =\frac{1}{2}(1-sgn(x))\Psi _{+}^{\ast }\left( -x\right) +\frac{1}{2%
}(1+sgn(x))\Psi _{-}^{\ast }\left( -x\right) \;.
\end{equation*}

Imposing $\mathcal{PT}$ $\Psi _{int.}\left( x\right) =e^{i\varphi }\Psi
_{int.}\left( x\right) $, with $\varphi $ a real phase, yields%
\begin{equation*}
\Psi _{\pm }^{\ast }\left( -x\right) =e^{i\varphi }\Psi _{\mp }\left(
x\right) \;.
\end{equation*}

Remembering the behaviour of $k_{\pm }$, $C_{\pm }$ and $D_{\pm }$ under
complex conjugation, we obtain the following constraints on $A_{\pm }$ and $%
B_{\pm }$%
\begin{equation}
\begin{array}{c}
A_{\pm }^{\ast }=e^{i\varphi }A_{\mp }\;, \\ 
B_{\pm }^{\ast }=e^{i\varphi }B_{\mp }\;.%
\end{array}
\label{PT_constr}
\end{equation}

For a progressive wave ($A_{-}=1$, $B_{+}=0$), this is equivalent to $%
T_{L\rightarrow R}=A_{+}=e^{-i\varphi _{-}}$ and $R_{L\rightarrow R}=B_{-}=0$%
, while, for a regressive wave ($\widetilde{A}_{-}=0$, $\ \widetilde{B}%
_{+}=1 $), one obtains $T_{R\rightarrow L}=\widetilde{B}_{\_}=e^{-i%
\widetilde{\varphi }}$ and $R_{R\rightarrow L}=\widetilde{A}_{+}=0$. In
other words, the potentials are reflectionless and conserve unitarity, since
the transmission coefficients have unit modulus.

In the non-relativistic limit, $|C_{\pm }|$, $|D_{\pm }|\ll 1$ and the lower
components of Dirac spinors are negligible with respect to the higher ones.
Non-vanishing potentials at $x\rightarrow \pm \infty $ only affect
asymptotic momenta $k_{\pm }$ and the preceding discussion and its
conclusions remain valid, thus generalizing the case of short-range
potentials treated in Ref.\cite{CDV07}.

It is worthwhile to point out that potentials that behave asymptotically
like $\mathcal{PT}$-symmetric step functions ($P_{+}=-P_{-}^{\ast }$ and so
on) may admit asymptotic wave functions that are eigenstates of $\mathcal{PT}
$, unlike the step functions themselves, which are not reflectionless,
because the asymptotic behaviour of wave functions is determined by the
behaviour of the potentials in their whole domain.

In the following sections, we specialize the general interaction of Eq. (\ref%
{gen_Dir_eq}) to scalar-plus-vector, pseudoscalar and scalar potentials and,
for each type of potential, work out some examples in detail.

\section{Scalar-plus-vector potentials}

\bigskip In the present section, we specialize Eq. (\ref{gen_Dir_eq}) to a
scalar-plus-vector potential with the same $x$ dependence: $S\left( x\right)
=c_{S}f\left( x\right) $, $V\left( x\right) =c_{V}f\left( x\right) $, with $%
c_{S}$ and $c_{V}$ real coupling constants. 
\begin{equation}
\left[ \alpha _{x}p_{x}+\beta m+\left( c_{S}\beta +c_{V}\right) f\left(
x\right) \right] \Psi \left( x\right) =E\Psi \left( x\right) \;,
\label{scal_vec_eq}
\end{equation}

\ We find it convenient to adopt the Dirac representation $\alpha
_{x}=\sigma _{x}$ and $\beta =\sigma _{z}$. The Dirac equation (\ref%
{scal_vec_eq}) satisfied by the spinor $\Psi \left( x\right) =\left( 
\begin{array}{c}
\psi _{1}\left( x\right) \\ 
\psi _{2}\left( x\right)%
\end{array}%
\right) $ is thus written explicitly in matrix form%
\begin{equation}
\left( 
\begin{array}{cc}
m-E+\left( c_{V}+c_{S}\right) f\left( x\right) & -i\frac{d}{dx} \\ 
-i\frac{d}{dx} & -m-E+\left( c_{V}-c_{S}\right) f\left( x\right)%
\end{array}%
\right) \left( 
\begin{array}{c}
\psi _{1}\left( x\right) \\ 
\psi _{2}\left( x\right)%
\end{array}%
\right) =\left( 
\begin{array}{c}
0 \\ 
0%
\end{array}%
\right) \;,  \label{mat_eq}
\end{equation}%
which reduces to a system of two first-order equations for the unknown
spinor components $\psi _{1}\left( x\right) $ and $\psi _{2}\left( x\right) $
. In order to obtain analytic solutions, we limit ourselves to the
particular cases $c_{V}=c_{S}$ and $c_{V}=-c_{S}$, which correspond to spin
symmetry and pseudo-spin symmetry in three space dimensions\cite{Gi05}. Let
us consider the case $c_{V}=c_{S}=c^{\prime }$ first. It is easy to see that
the two equations obtained from formula (\ref{mat_eq}) reduce to the simple
system%
\begin{equation}
\left\{ 
\begin{array}{c}
-\frac{d^{2}}{dx^{2}}\psi _{1}\left( x\right) +2c^{\prime }\left( E+m\right)
f\left( x\right) \psi _{1}\left( x\right) =\left( E^{2}-m^{2}\right) \psi
_{1}\left( x\right) \equiv k^{2}\psi _{1}\left( x\right) \\ 
\psi _{2}\left( x\right) =-\frac{i}{E+m}\frac{d}{dx}\psi _{1}\left( x\right)%
\end{array}%
\right. \;.  \label{sys_1}
\end{equation}

Here, the equation satisfied by $\psi _{1}\left( x\right) $ is Schr\"{o}%
dinger-like, with the same $\mathcal{PT}$-symmetric form $f\left( x\right) $
as the original Dirac equation and an energy-dependent potential strength $%
s^{\prime }\left( E\right) =2c^{\prime }\left( E+m\right) $ and $\psi
_{2}\left( x\right) $ is obtained by deriving $\psi _{1}\left( x\right) $
with respect to $x$. On the r.h.s. of the first equation, $k^{2}>0$ for
scattering states, while for bound states, $k^{2}<0$ implies an imaginary
value of $k$, corresponding to poles of the transmission coefficient. In the
limiting case $k=0$, both normalizable bound states and non-normalizable
half-bound states\cite{KD02}, corresponding to transmission resonances, are
possible, depending on the potentials under consideration.

When $c_{V}=-c_{S}=c^{\prime \prime }$, $\psi _{1}$ and $\psi _{2}$ exchange
their role, since $\psi _{2}$ now satisfies a Schr\"{o}dinger-like equation
with the original $f\left( x\right) $ and the energy-dependent strength $%
s^{\prime \prime }\left( E\right) =2c^{\prime \prime }\left( E-m\right) $,
while $\psi _{1}$ is proportional to the space derivative of $\psi _{2}$.%
\begin{equation}
\left\{ 
\begin{array}{c}
-\frac{d^{2}}{dx^{2}}\psi _{2}\left( x\right) +2c^{\prime \prime }\left(
E-m\right) f\left( x\right) \psi _{2}\left( x\right) =\left(
E^{2}-m^{2}\right) \psi _{2}\left( x\right) \equiv k^{2}\psi _{2}\left(
x\right) \\ 
\psi _{1}\left( x\right) =-\frac{i}{E-m}\frac{d}{dx}\psi _{2}\left( x\right)%
\end{array}%
\right.  \label{sys_2}
\end{equation}

Energy dependence of the coupling strengths in Eqs. (\ref{sys_1}-\ref{sys_2}%
) may affect the reflection properties of a $\mathcal{PT}$-symmetric
potential. An example of this general behaviour is provided by the
hyperbolic Scarf potential with integer coupling constants $l$ and $n$%
\begin{equation}
f\left( x\right) =-\frac{l^{2}+n\left( n+1\right) }{2m}\frac{1}{\cosh ^{2}x}+%
\frac{il\left( 2n+1\right) }{2m}\frac{\sinh x}{\cosh ^{2}x}\;,
\label{hyper_scarf}
\end{equation}%
which is known to be reflectionless in the Schr\H{o}dinger equation\cite%
{CDV07} (note that the quoted reference uses units $2m=1$, as is common in
non-relativistic quantum mechanics). When inserted in the Dirac equation, it
gives rise to an equivalent Schr\H{o}dinger-like equation (\ref{sys_1})
where the potential maintains the same shape, but is no more reflectionless,
because of the energy dependence of the coupling strengths.

On the contrary, if $f\left( x\right) $ exhibits an exact $\mathcal{PT}$
symmetry in the Schr\"{o}dinger problem, it maintains it in the Dirac
problem with the appropriate superposition of vector and scalar components,
provided it is not connected with a particular value of the coupling
strength, $s^{\prime }$ or $s^{\prime \prime }$, which becomes a function of 
$E$.

A notable example is provided by the $\mathcal{PT}$-symmetric potential%
\begin{equation}
f\left( x\right) =\frac{1}{\left( x+i\epsilon \right) ^{2}}\;,
\label{Centrif}
\end{equation}%
where $\epsilon $ is an arbitrary real number, regularizing $f$ at $x=0$,
which is a well-known example of reflectionless potential in the Schr\"{o}%
dinger case\cite{CDV07}. Let us consider the case $c_{V}=c_{S}=c^{\prime }$
first, so that the equation (\ref{sys_1}) satisfied by $\psi _{1}$ reads 
\begin{equation}
-\frac{d^{2}}{dx^{2}}\psi _{1}\left( x\right) +\frac{2c^{\prime }\left(
E+m\right) }{\left( x+i\epsilon \right) ^{2}}\psi _{1}\left( x\right)
=\left( E^{2}-m^{2}\right) \psi _{1}\left( x\right) \equiv k^{2}\psi
_{1}\left( x\right)  \label{psi_1}
\end{equation}%
for scattering states ($E^{2}>m^{2}$). The above equation is Schr\"{o}%
dinger-like and is quickly solved by introducing the complex variable $%
z=k\left( x+i\epsilon \right) $ and factorizing $\psi _{1}\left( z\right)
=z^{1/2}\varphi \left( z\right) $: in fact, the equation satisfied by $%
\varphi $%
\begin{equation}
z^{2}\frac{d^{2}}{dz^{2}}\varphi \left( z\right) +z\frac{d}{dz}\varphi
\left( z\right) +\left[ z^{2}-2c^{\prime }(m+E)-\frac{1}{4}\right] \varphi
\left( z\right) =0  \label{Bessel_eq}
\end{equation}%
is a Bessel equation of index $\nu ^{2}=2c^{\prime }(m+E)+\frac{1}{4}$. Note
that $\nu $ is imaginary when $2c^{\prime }(m+E)+\frac{1}{4}<0$, which can
happen, for instance, for positive $c^{\prime }$ and large negative $E$, or
viceversa.

Two linearly independent solutions of Eq. (\ref{Bessel_eq}) with asymptotic
behaviour appropriate to scattering states are the Hankel functions of first
and second type, $H_{\nu }^{\left( 1\right) }\left( z\right) $ and $H_{\nu
}^{\left( 2\right) }\left( z\right) $, respectively%
\begin{equation}
\begin{array}{c}
\lim_{|z|\rightarrow \infty }H_{\nu }^{\left( 1\right) }\left( z\right)
=\left( \frac{2}{\pi z}\right) ^{1/2}\exp \left[ i\left( z-\frac{\pi }{2}\nu
-\frac{\pi }{4}\right) \right] \;, \\ 
\lim_{|z|\rightarrow \infty }H_{\nu }^{\left( 2\right) }\left( z\right)
=\left( \frac{2}{\pi z}\right) ^{1/2}\exp \left[ -i\left( z-\frac{\pi }{2}%
\nu -\frac{\pi }{4}\right) \right] \;,%
\end{array}
\label{Hankel}
\end{equation}%
valid for $\Re \nu >-1/2$, $|\arg z|<\pi $, this latter condition being
ensured by the non-zero imaginary part of $z$, \textit{i.e. }$\Im
z=k\epsilon $. According to formulae (\ref{sys_1}), the corresponding
linearly independent solutions of the Dirac equation are%
\begin{eqnarray*}
\Psi ^{\left( k\right) }\left( x\right) &\equiv &\left( 
\begin{array}{c}
\psi _{1}^{\left( k\right) }\left( x\right) \\ 
-\frac{i}{E+m}\frac{d}{dx}\psi _{1}^{\left( k\right) }\left( x\right)%
\end{array}%
\right) =z^{1/2}\left( 
\begin{array}{c}
H_{\nu }^{\left( k\right) }\left( z\right) \\ 
-i\lambda \left( \frac{d}{dz}H_{\nu }^{\left( k\right) }\left( z\right) +%
\frac{1}{2z}H_{\nu }^{\left( k\right) }\left( z\right) \right)%
\end{array}%
\right) \\
&=&z^{1/2}\left( 
\begin{array}{c}
H_{\nu }^{\left( k\right) }\left( z\right) \\ 
-i\lambda \left( H_{\nu -1}^{\left( k\right) }\left( z\right) +\frac{1-2\nu 
}{2z}H_{\nu }^{\left( k\right) }\left( z\right) \right)%
\end{array}%
\right) \;,\;\;\left( k=1,2\right)
\end{eqnarray*}%
where $\lambda \equiv \frac{k}{E+m}$. In order to obtain the final form of
the r.h.s., use has been made of the relation $\frac{d}{dz}H_{\nu }^{\left(
k\right) }\left( z\right) =H_{\nu -1}^{\left( k\right) }\left( z\right) -%
\frac{\nu }{z}H_{\nu }^{\left( k\right) }\left( z\right) $. The asymptotic
behaviour of the Dirac spinors is%
\begin{equation*}
\lim_{x\rightarrow \pm \infty }\Psi ^{\left( k\right) }\left( x\right)
=\lim_{|z|\rightarrow \infty }z^{1/2}\left( 
\begin{array}{c}
H_{\nu }^{\left( k\right) }\left( z\right) \\ 
-i\lambda H_{\nu -1}^{\left( k\right) }\left( z\right)%
\end{array}%
\right) \;,
\end{equation*}%
or, more explicitly, using formulae (\ref{Hankel})%
\begin{equation}
\lim_{x\rightarrow \pm \infty }\Psi ^{\left( 1\right) }\left( x\right)
=\left( \frac{2}{\pi }\right) ^{1/2}\left( 
\begin{array}{c}
1 \\ 
\lambda%
\end{array}%
\right) \exp \left( ikx-k\epsilon -i\frac{\pi }{2}\nu -i\frac{\pi }{4}\right)
\label{Psi_1}
\end{equation}%
and 
\begin{equation}
\lim_{x\rightarrow \pm \infty }\Psi ^{\left( 2\right) }\left( x\right)
=\left( \frac{2}{\pi }\right) ^{1/2}\left( 
\begin{array}{c}
1 \\ 
-\lambda%
\end{array}%
\right) \exp \left( -ikx+k\epsilon +i\frac{\pi }{2}\nu +i\frac{\pi }{4}%
\right) \;.  \label{Psi_2}
\end{equation}

Formulae (\ref{Psi_1}-\ref{Psi_2}) are particular cases of the asymptotic
formulae of Section 2, whose constants now are%
\begin{equation*}
a_{1-}=a_{1+}=\left( \frac{2}{\pi }\right) ^{1/2}\exp \left( -k\epsilon -i%
\frac{\pi }{2}\nu -i\frac{\pi }{4}\right) \;,
\end{equation*}%
\begin{equation*}
b_{1-}=b_{1+}=0\;,
\end{equation*}%
\begin{equation*}
a_{2-}=a_{2+}=0\;,
\end{equation*}%
\begin{equation*}
b_{2-}=b_{2+}=\left( \frac{2}{\pi }\right) ^{1/2}\exp \left( k\epsilon +i%
\frac{\pi }{2}\nu +i\frac{\pi }{4}\right) \;,
\end{equation*}%
so that formulae (\ref{TR_LR}-\ref{TR_RL}) immediately yield%
\begin{equation*}
T_{L\rightarrow R}=T_{R\rightarrow L}=1\;,\;R_{L\rightarrow
R}=R_{R\rightarrow L}=0\;.
\end{equation*}

Of course, potential (\ref{Centrif}) does not sustain bound states with $%
k\neq 0$, \ because the transmission coefficients are independent of $k$ and
cannot have poles in $k$, or $E$. \ When $k=0$, a non-trivial solution can
exist at $E=m$: in this case, Eq. (\ref{psi_1}) reduces to 
\begin{equation}
-\frac{d^{2}}{dx^{2}}\psi _{1}\left( x\right) +\frac{4c^{\prime }m}{\left(
x+i\epsilon \right) ^{2}}\psi _{1}\left( x\right) =0\;.  \label{psi_11}
\end{equation}

The solution to Eq. (\ref{psi_11}) can be searched for in the form of a
power, $\left( x+i\epsilon \right) ^{\gamma }$, thus leading to an algebraic
equation for $\gamma $%
\begin{equation*}
\gamma ^{2}-\gamma -4c^{\prime }m=0\;,
\end{equation*}%
whose solutions are 
\begin{equation*}
\gamma _{1,2}=\frac{1\pm \left( 1+16c^{\prime }m\right) ^{1/2}}{2}\;.
\end{equation*}

Depending on whether $1+16c^{\prime }m\gtrless 0$, the two roots are either
real or complex conjugate: in both cases, the general solution to Eq. (\ref%
{psi_11}) can be put in the form%
\begin{equation*}
\psi _{1}\left( x\right) =\alpha _{1}\left( x+i\epsilon \right) ^{\gamma
_{1}}+\alpha _{2}\left( x+i\epsilon \right) ^{\gamma _{2}}\;\;,
\end{equation*}%
where $\alpha _{i}$ $\left( i=1,2\right) $ are to be fixed on boundary
conditions. It is easy to understand that a normalizable solution, \textit{%
i.e.} a bound state, can exist only when $1+16c^{\prime }m>1$, or $c^{\prime
}>0$, by choosing $\alpha _{1}=0$. In this case, the solution for $\psi
_{1}\left( x\right) $ reads 
\begin{equation}
\psi _{1}\left( x\right) =\alpha _{2}\left( x+i\epsilon \right) ^{\frac{%
1-\beta }{2}}\;.  \label{psi_1_sol}
\end{equation}

Here, $\beta =\sqrt{1+16c^{\prime }m}>1$ and $\alpha _{2}$ can be determined
by normalization of the complete Dirac spinor 
\begin{equation}
\int\limits_{-\infty }^{+\infty }dx\left( 
\begin{array}{cc}
\psi _{1}^{\ast }\left( x\right) & \frac{i}{2m}\frac{d}{dx}\psi _{1}^{\ast
}\left( x\right)%
\end{array}%
\right) \left( 
\begin{array}{c}
\psi _{1}\left( x\right) \\ 
-\frac{i}{2m}\frac{d}{dx}\psi _{1}\left( x\right)%
\end{array}%
\right) =1\;.  \label{norm}
\end{equation}

Both integrals in formula (\ref{norm}) can be computed analytically in terms
of asymptotic expansions of the hypergeometric function, $F\left(
A,B,C;z\right) $, since they can be reduced to the integral representation 
\begin{equation*}
\int \left( 1+y^{2}\right) ^{-a}dy=yF\left( \frac{1}{2},a,\frac{3}{2}%
;-y^{2}\right) +const.\;\;\left( \Re a>\frac{1}{2}\right)
\end{equation*}%
yielding%
\begin{equation*}
\int_{-\infty }^{+\infty }\left( 1+y^{2}\right) ^{-a}dy\simeq \sqrt{\pi }%
\frac{\Gamma \left( a-\frac{1}{2}\right) }{\Gamma \left( a\right) }\;,
\end{equation*}%
where $\Gamma \left( a\right) $ is the Euler gamma function.

The final result for the normalization constant is%
\begin{equation*}
|\alpha _{2}|^{2}=\frac{\epsilon ^{\beta }}{\sqrt{\pi }\left[ \epsilon ^{2}%
\frac{\Gamma \left( \frac{\beta }{2}-1\right) }{\Gamma \left( \frac{\beta -1%
}{2}\right) }+\frac{1}{4m^{2}}\left( \frac{1-\beta }{2}\right) ^{2}\frac{%
\Gamma \left( \frac{\beta }{2}\right) }{\Gamma \left( \frac{\beta +1}{2}%
\right) }\right] }\;.
\end{equation*}

For the sake of completeness, we mention the case of the double root, $%
\gamma =1/2$, of the characteristic equation, occurring when $c^{\prime }=-%
\frac{1}{16m}$: the general solution to Eq. (\ref{psi_11}) for this case can
be put in the form%
\begin{equation*}
\psi _{1}\left( x\right) =\alpha _{1}\left( x+i\epsilon \right)
^{1/2}+\alpha _{2}\left( x+i\epsilon \right) ^{1/2}\ln \left( x+i\epsilon
\right)
\end{equation*}%
and is not normalizable.

The case $c_{V}=-c_{S}=c^{\prime \prime }$ can be treated in a similar way
starting from Eqs. (\ref{sys_2}): the solution of the Schr\"{o}dinger-like
equation for $\psi _{2}\left( x\right) $ is obtained in the same way as
before and the bound state with $k=0$ now appears at $E=-m$.

\section{Pseudoscalar potentials}

The Dirac equation with a pseudoscalar potential, $P\left( x\right) \equiv
c_{P}f\left( x\right) $, and $c_{P}$ a real coupling constant, reads%
\begin{equation}
\left[ \alpha _{x}p_{x}+\beta m+ic_{P}\alpha _{x}\beta f\left( x\right) %
\right] \Psi \left( x\right) =E\Psi \left( x\right) \;.  \label{ps_Dirac_eq}
\end{equation}

In the Dirac representation, $i\alpha _{x}\beta =i\sigma _{x}\sigma
_{z}=\sigma _{y}=\left( 
\begin{array}{cc}
0 & -i \\ 
i & 0%
\end{array}%
\right) $. $\mathcal{PT}$ invariance of the interaction term implies $%
f^{\ast }\left( -x\right) =-f\left( x\right) $. After expressing Eq. (\ref%
{ps_Dirac_eq}) as a system of coupled equations in the two components, $\psi
_{1}\left( x\right) $ and $\psi _{2}\left( x\right) $, of the Dirac spinor, $%
\Psi \left( x\right) $,%
\begin{equation}
\left( 
\begin{array}{cc}
m-E & -i\frac{d}{dx}-iP\left( x\right) \\ 
-i\frac{d}{dx}+iP\left( x\right) & -m-E%
\end{array}%
\right) \left( 
\begin{array}{c}
\psi _{1}\left( x\right) \\ 
\psi _{2}\left( x\right)%
\end{array}%
\right) =\left( 
\begin{array}{c}
0 \\ 
0%
\end{array}%
\right) \;,  \label{ps_sys}
\end{equation}%
it is almost immediate to derive the two decoupled Schr\H{o}dinger-like
equations satisfied by $\psi _{1}$ and $\psi _{2}$%
\begin{equation}
\left( -\frac{1}{2m}\frac{d^{2}}{dx^{2}}+U_{j}(x)\right) \psi _{j}\left(
x\right) =\frac{E^{2}-m^{2}}{2m}\psi _{j}\left( x\right) \equiv \varepsilon
\psi _{j}\left( x\right) \;,\;\left( j=1,2\right)  \label{susy_p_eq}
\end{equation}%
where $U_{1}\left( x\right) \equiv \frac{1}{2m}(P^{2}\left( x\right) +\frac{d%
}{dx}P\left( x\right) )$, $U_{2}\left( x\right) \equiv \frac{1}{2m}%
(P^{2}\left( x\right) -\frac{d}{dx}P\left( x\right) )$. As already shown in
Ref.\cite{SR05}, the two $\mathcal{PT}$-symmetric Hamiltonians%
\begin{equation}
H_{j}\equiv -\frac{1}{2m}\frac{d^{2}}{dx^{2}}+U_{j}(x),\;\left( j=1,2\right)
\label{H_1,2}
\end{equation}%
constitute the Bose sector of a non-Hermitian representation of an $sl(1|1)$
superalgebra\cite{CKS95}. The corresponding super-Hamiltonian is%
\begin{equation}
\mathcal{H=}\left( 
\begin{array}{cc}
H_{1} & 0 \\ 
0 & H_{2}%
\end{array}%
\right) \;,  \label{bose_sector}
\end{equation}%
while the differential operators%
\begin{equation}
L\equiv \frac{1}{\sqrt{2m}}\left( \frac{d}{dx}+P\left( x\right) \right)
\;,\;M\equiv \frac{1}{\sqrt{2m}}\left( -\frac{d}{dx}+P\left( x\right) \right)
\label{L_M}
\end{equation}%
forming the two partner Hamiltonians $H_{1}=LM$ and $H_{2}=ML$, give also
rise to the "supercharges"%
\begin{equation}
Q_{1}\equiv \left( 
\begin{array}{cc}
0 & L \\ 
0 & 0%
\end{array}%
\right) \;,\;Q_{2}\equiv \left( 
\begin{array}{cc}
0 & 0 \\ 
M & 0%
\end{array}%
\right) \;,  \label{fermi_sector}
\end{equation}%
which form the Fermi sector, since $\mathcal{H}$, $Q_{1}$ and $Q_{2}$ are
closed under the following set of commutation and anticommutation relations 
\begin{equation*}
\left\{ Q_{1},Q_{2}\right\} _{+}=\mathcal{H\;},\;\left[ Q_{1},\mathcal{H}%
\right] _{-}=\left[ Q_{2},\mathcal{H}\right] _{-}=0\;.
\end{equation*}

It is worthwhile to point out that, differently from the Hermitian case, $%
Q_{2}$ is not the Hermitian adjoint of $Q_{1}$, but the following relations
hold%
\begin{equation*}
M=(\mathcal{PT)}L\left( \mathcal{PT}\right) ^{-1}=\mathcal{P}L^{\dag }%
\mathcal{P}^{-1}\;,\;Q_{2}=\mathcal{P}Q_{1}^{\dag }\mathcal{P}^{-1}\;.
\end{equation*}

One thus speaks, in this case, of a $\mathcal{P}$-pseudo-supersymmetry\cite%
{Mo02}. Note that $P\left( x\right) $ plays the role of a superpotential.
The fact that the supercharges (\ref{fermi_sector}) are not the Hermitian
conjugates of each other gives rise to a richer structure of supersymmetric
systems compared with Hermitian theories\cite{Ba09}.

As is known, the discrete spectra and the scattering properties of the
partner Hamiltonians are connected by supersymmetry, so that it is
sufficient to compute bound states and scattering states of one partner only.

In particular, it is not difficult to prove the relations connecting the
transmission and reflection coefficients of the two partners, in terms of
the asymptotic limits of the superpotential, $P_{\pm }=\lim_{x\rightarrow
\pm \infty }P\left( x\right) $ and the corresponding asymptotic momenta, $%
k_{\pm }^{2}=E^{2}-m^{2}-P_{\pm }^{2}$: let $T_{L\rightarrow R}^{\left(
j\right) }$ ($R_{L\rightarrow R}^{\left( j\right) })$ be the transmission
(reflection) coefficient of a progressive wave in presence of potential $%
U^{\left( j\right) }$ ($j=1,2$) : we easily obtain%
\begin{equation}
\begin{array}{c}
R_{L\rightarrow R}^{\left( 1\right) }=\frac{-ik_{-}+P_{-}}{ik_{-}+P_{-}}%
R_{L\rightarrow R}^{\left( 2\right) } \\ 
T_{L\rightarrow R}^{\left( 1\right) }=\frac{ik_{+}+P_{+}}{ik_{-}+P_{-}}%
T_{L\rightarrow R}^{\left( 2\right) }%
\end{array}%
\;.  \label{T12_R_12}
\end{equation}

Similar relations connect the transmission and reflection coefficients of a
regressive wave%
\begin{equation}
\begin{array}{c}
R_{R\rightarrow L}^{\left( 1\right) }=\frac{ik_{+}+P_{+}}{-ik_{+}+P_{+}}%
R_{R\rightarrow L}^{\left( 2\right) } \\ 
T_{R\rightarrow L}^{\left( 1\right) }=\frac{-ik_{-}+P_{-}}{-ik_{+}+P_{+}}%
T_{R\rightarrow L}^{\left( 2\right) }%
\end{array}
\label{T12_R_12_b}
\end{equation}

Derivation of formulae (\ref{T12_R_12}-\ref{T12_R_12_b}) is given in
Appendix , for the sake of completeness.

It is obvious that if one of the two partners is reflectionless, so is the
other and, consequently, the superpotential in the Dirac equation.

In Ref.\cite{NT98}, the Kay-Moses method of constructing reflectionless
potentials was extended to real symmetric pseudoscalar potentials in the
Dirac equation. The examples worked out in the quoted reference can be made $%
\mathcal{PT}$-symmetric by applying an imaginary shift to the space
coordinate, $x\rightarrow x+i\epsilon $. Thus, the following superpotential,
with $c_{P}=1$ for simplicity's sake%
\begin{equation}
f\left( x\right) =\tanh \left( x+i\epsilon \right) +\frac{\lambda ^{2}-1}{%
\tanh \left( x+i\epsilon \right) -\lambda \coth \left( \lambda \left(
x+i\epsilon \right) \right) }\;,  \label{nogami_toyama}
\end{equation}%
and $\lambda \geq 1$, generates the following supersymmetry partners%
\begin{equation*}
U_{1}\left( x\right) =\frac{1}{2m}\left[ \lambda ^{2}-\frac{2\left( \lambda
^{2}-1\right) \left( \lambda ^{2}\cosh ^{2}\left( x+i\epsilon \right) +\sinh
^{2}\left( \lambda \left( x+i\epsilon \right) \right) \right) }{\left(
\lambda \cosh \left( x+i\epsilon \right) \cosh \left( \lambda \left(
x+i\epsilon \right) \right) -\sinh \left( x+i\epsilon \right) \sinh \left(
\lambda \left( x+i\epsilon \right) \right) \right) ^{2}}\right] \;,
\end{equation*}%
and 
\begin{equation}
U_{2}\left( x\right) =\frac{1}{2m}\left( \lambda ^{2}-\frac{2}{\cosh
^{2}\left( x+i\epsilon \right) }\right) \;.  \label{poescl_teller}
\end{equation}

A part from the constant term $\lambda ^{2}/\left( 2m\right) $, which enters
in the definition of the asymptotic momentum, $U_{2}\left( x\right) $ is a
reflectionless P\H{o}schl-Teller potential\cite{CDV07} and $U_{1}\left(
x\right) $ is necessarily reflectionless, too. It is worthwhile to point out
that our definitions of $U_{1}$ and $U_{2}$ are exchanged with respect to
Ref.\cite{NT98}, but in agreement with Ref.\cite{SR05}. Transmission and
reflection coefficients for potential (\ref{poescl_teller}) can be
immediately written down from the corresponding formulae of the more general
hyperbolic Scarf potential obtained in Ref.\cite{CDV07}, after observing
that $f_{\pm }=\mp \lambda $ and the asymptotic momenta are $k_{\pm }=\sqrt{%
E^{2}-m^{2}-V_{\pm }^{2}}=\sqrt{E^{2}-m^{2}-\lambda ^{2}}\equiv k$%
\begin{equation}
\begin{array}{c}
R_{L\rightarrow R}^{\left( 2\right) }=R_{R\rightarrow L}^{\left( 2\right)
}=0\;, \\ 
T_{L\rightarrow R}^{\left( 2\right) }=T_{R\rightarrow L}^{\left( 2\right) }=-%
\frac{1-ik}{1+ik}\;.%
\end{array}
\label{RT_PT}
\end{equation}

A real $k$ is a necessary condition for $\left\vert T_{L\rightarrow
R}^{\left( 2\right) }\right\vert =\left\vert T_{R\rightarrow L}^{\left(
2\right) }\right\vert =1$, equivalent to exact $\mathcal{PT}$ symmetry of
the asymptotic wave functions.

Formulae (\ref{T12_R_12}-\ref{T12_R_12_b}) immediately yield the
corresponding coefficients for potential $U_{1}$%
\begin{equation*}
\begin{array}{c}
R_{L\rightarrow R}^{\left( 1\right) }=R_{R\rightarrow L}^{\left( 1\right)
}=0\;, \\ 
T_{L\rightarrow R}^{\left( 1\right) }=T_{R\rightarrow L}^{\left( 1\right) }=%
\frac{\lambda -ik}{\lambda +ik}\frac{1-ik}{1+ik}\;.%
\end{array}%
\end{equation*}

As far as bound states are concerned, it is well known\cite{CKS95} that $%
U_{2}\left( x\right) $ admits only one bound state with eigenvalue $%
\varepsilon =\frac{\lambda ^{2}-1}{2m}$ ( or $E^{2}=m^{2}+\lambda ^{2}-1$),
and the corresponding wave function is 
\begin{equation}
\psi _{2}\left( x\right) =\frac{N}{\cosh \left( x+i\epsilon \right) }\;,
\label{psi2}
\end{equation}%
with the constant $N$ to be determined from normalization of the complete
Dirac spinor.

Since $H_{2}\psi _{2}\left( x\right) =ML\psi _{2}\left( x\right) =\frac{%
\lambda ^{2}-1}{2m}\psi _{2}\left( x\right) $, we have that $LH_{2}\psi
_{2}\left( x\right) =LML\psi _{2}\left( x\right) =H_{1}L\psi _{2}\left(
x\right) =\frac{\lambda ^{2}-1}{2m}L\psi _{2}\left( x\right) $. Therefore $%
\psi _{1}\left( x\right) =c_{0}L\psi _{2}\left( x\right) $, with $c_{0}$ a
normalization constant, is eigenfunction of $H_{1}$ with eigenvalue $\frac{%
\lambda ^{2}-1}{2m}$ and corrresponds to the first component of the Dirac
spinor. From the definition of the differential operator $L$ and from the
first equation (\ref{ps_sys}) we get $c_{0}=i\frac{\sqrt{2m}}{m-E}$ and 
\begin{eqnarray}
\psi _{1}\left( x\right) &=&\frac{i}{m-E}\left( \frac{d}{dx}+f\left(
x\right) \right) \frac{N}{\cosh \left( x+i\epsilon \right) }  \label{psi1} \\
&=&\left( \frac{iN}{m-E}\right) \frac{\lambda ^{2}-1}{\sinh \left(
x+i\epsilon \right) -\lambda \cosh \left( x+i\epsilon \right) \coth \left(
\lambda \left( x+i\epsilon \right) \right) }  \notag
\end{eqnarray}%
with $\int\limits_{-\infty }^{+\infty }\Psi ^{\dag }\left( x\right) \Psi
\left( x\right) dx=\int\limits_{-\infty }^{+\infty }\left( \left\vert \psi
_{1}\left( x\right) \right\vert ^{2}+\left\vert \psi _{2}\left( x\right)
\right\vert ^{2}\right) dx=1$.

$\psi _{1}\left( x\right) $ from formula (\ref{psi1}) has a node at $%
x=-i\epsilon $ and is not the ground state of $H_{1}$. Conversely, if a
non-trivial normalizable solution of the equation $M\overline{\psi }%
_{1}\left( x\right) =0$ exists, then $H_{1}\overline{\psi }_{1}\left(
x\right) =LM\overline{\psi }_{1}\left( x\right) =0$, and $\overline{\psi }%
_{1}\left( x\right) $ is eigenstate of $H_{1}$ with eigenvalue $\varepsilon
=0$. In this case, $\overline{\psi }_{1}\left( x\right) $ cannot be written
as $c_{0}L\overline{\psi }_{2}\left( x\right) $, with $\overline{\psi }%
_{2}\left( x\right) $ a non-trivial normalizable function, since, otherwise,
we would have $ML\overline{\psi }_{2}\left( x\right) =H_{2}\overline{\psi }%
_{2}\left( x\right) =0$ and $\overline{\psi }_{2}\left( x\right) $ would be
an eigenstate of $H_{2}$ with eigenvalue $\varepsilon =0$, which is
impossible, because $H_{2}\overline{\psi }_{2}\left( x\right) =0$ admits
only the trivial solution $\overline{\psi }_{2}\left( x\right) =0$.

The equation%
\begin{equation}
M\overline{\psi }_{1}\left( x\right) =\frac{1}{\sqrt{2m}}\left( -\frac{d}{dx}%
+f\left( x\right) \right) \overline{\psi }_{1}\left( x\right) =0\;,
\label{M_psi1}
\end{equation}
is satisfied by%
\begin{equation*}
\overline{\psi }_{1}\left( x\right) =\exp \left( \int\limits^{x}f\left(
x^{\prime }\right) dx^{\prime }\right) \;.
\end{equation*}

\bigskip Note that the condition $\lim_{x\rightarrow \pm \infty }f\left(
x\right) \equiv f_{\pm }=\mp \lambda $, with $\lambda >1$, yields $%
\lim_{x\rightarrow \pm \infty }\overline{\psi }_{1}\left( x\right) =0$.

In our case%
\begin{equation*}
\int^{x}f\left( x\right) dx=\ln \left( \cosh x\right) -\ln \left( 2\lambda
\cosh x\cosh \left( \lambda x\right) -2\sinh x\sinh \left( \lambda x\right)
\right) +\ln \overline{N}\;,
\end{equation*}%
so that%
\begin{equation}
\overline{\psi }_{1}^{.}\left( x\right) =\frac{\overline{N}}{\lambda \cosh
\left( \lambda (x+i\epsilon )\right) -\tanh \left( x+i\epsilon \right) \sinh
\left( \lambda \left( x+i\epsilon \right) \right) }\;,  \label{psi1_gs}
\end{equation}%
where $\overline{N}$ is to be determined from normalization. The second
component of the Dirac spinor, $\overline{\psi }_{2}^{.}\left( x\right) $,
is solution of the equation $H_{2}\overline{\psi }_{2}\left( x\right) =ML%
\overline{\psi }_{2}\left( x\right) =0\Rightarrow L\overline{\psi }%
_{2}\left( x\right) =0$, which admits only the trivial solution, $\overline{%
\psi }_{2}\left( x\right) =0$. $\overline{N}$ is thus determined from the
condition $\int\limits_{-\infty }^{+\infty }\Psi ^{\dag }\left( x\right)
\Psi \left( x\right) dx=\int\limits_{-\infty }^{+\infty }\left\vert 
\overline{\psi }_{1}\left( x\right) \right\vert ^{2}dx=1$.

In this case, the ground state of $H_{2}$ has the same energy, $\varepsilon =%
\frac{\lambda ^{2}-1}{2m}>0$, as the first excited state of $H_{1}$, whose
ground state has $\varepsilon =0$. This is an example of exact supersymmetry.

A second example of reflectionless pseudoscalar potential, whose bound
states were already studied in Ref.\cite{SR05}, is%
\begin{equation}
f\left( x\right) =n\tanh x+i\frac{l}{\cosh x}\;,  \label{super_scarf}
\end{equation}%
with integer constants $n$ and $l$. \ In fact, the two supersymmetric
partners from formula (\ref{H_1,2}) are%
\begin{equation*}
\begin{array}{c}
U_{1}\left( x\right) =\frac{1}{2m}\left( n^{2}-\frac{n\left( n-1\right)
+l^{2}}{\cosh ^{2}x}+il\left( 2n-1\right) \frac{\sinh x}{\cosh ^{2}x}\right)
\;, \\ 
U_{2}\left( x\right) =\frac{1}{2m}\left( n^{2}-\frac{n\left( n+1\right)
+l^{2}}{\cosh ^{2}x}+il\left( 2n+1\right) \frac{\sinh x}{\cosh ^{2}x}\right)
\;,%
\end{array}%
\end{equation*}%
which, a part from the constant term $\frac{n^{2}}{2m}$, are reflectionless
potentials of hyperbolic Scarf type (\ref{hyper_scarf}). In this case, $%
f_{\pm }=\pm n$ and $k_{\pm }=\sqrt{E^{2}-m^{2}-n^{2}}\equiv k$. Here again,
the transmission coefficients are given by Ref.\cite{CDV07}%
\begin{equation*}
T_{L\rightarrow R}^{\left( 2\right) }=T_{R\rightarrow L}^{\left( 2\right)
}=\left( -1\right) ^{n+l}\frac{\left( n-ik\right) \ldots \left( 1-ik\right)
\left( l-\frac{1}{2}-ik\right) \ldots \left( \frac{1}{2}-ik\right) }{\left(
n+ik\right) \ldots \left( 1+ik\right) \left( l-\frac{1}{2}+ik\right) \ldots
\left( \frac{1}{2}+ik\right) }
\end{equation*}%
for $n>1$ and%
\begin{equation*}
T_{L\rightarrow R}^{\left( 1\right) }=T_{R\rightarrow L}^{\left( 1\right)
}=\left( -1\right) ^{n+l-1}\frac{\left( n-1-ik\right) \ldots \left(
1-ik\right) \left( l-\frac{1}{2}-ik\right) \ldots \left( \frac{1}{2}%
-ik\right) }{\left( n-1+ik\right) \ldots \left( 1+ik\right) \left( l-\frac{1%
}{2}+ik\right) \ldots \left( \frac{1}{2}+ik\right) }\;
\end{equation*}%
for $n>2$.

Unitarity and asymptotic $\mathcal{PT}$ symmetry are conserved if $k$ is
real.

As for bound states, $U_{2}\left( x\right) $ admits $n$ of them, all with
real energies, and $U_{1}\left( x\right) $ has $n-1$ bound states at the
same energies of those of $U_{2}\left( x\right) $, excepted the ground state
of the latter. Here again, the pseudo-supersymmetry is exact.

\section{Scalar potentials}

When only a scalar potential, $S\left( x\right) =S^{\ast }\left( -x\right) $%
, is present, Eq. (\ref{gen_Dir_eq}) simplifies to%
\begin{equation}
\left[ \alpha _{x}p_{x}+\beta \left( m+S\left( x\right) \right) \right] \Psi
\left( x\right) =E\Psi \left( x\right) \;.  \label{scal_eq}
\end{equation}

In spite of its apparent greater simplicity, however, Eq. (\ref{scal_eq}) is
more difficult to solve than Eq. (\ref{scal_vec_eq}), including a vector
potential with the same $x$ dependence of the scalar potential and equal, or
opposite coupling strength, if one adopts the Dirac representation $\alpha
_{x}=\sigma _{x}$, $\beta =\sigma _{z}$, because the second order equation
satisfied by the first component, $\psi _{1}\left( x\right) $, of the Dirac
spinor, $\Psi \left( x\right) $, now contains also a first order derivative, 
$\frac{d}{dx}\psi _{1}\left( x\right) $, whose $x$-dependent coefficient is
negligible only at $x=\pm \infty $, provided $S\left( x\right) $ admits
constant limits, $\lim_{x\rightarrow \pm \infty }S\left( x\right) =S_{\pm }$%
, as assumed in Section (2) in deriving the general form of asymptotic wave
functions for arbitrary combinations of scalar, vector and pseudoscalar
potentials.

If we are interested in exact wave functions, including those corresponding
to bound states, it is more convenient to adopt a different representation
of Dirac matrices, $\alpha _{x}=\sigma _{y}$, $\beta =\sigma _{x}$, like in
Ref.\cite{SR05}. The main drawback of this choice is that the kinetic term
of the Dirac Hamiltonian, $\alpha _{x}p_{x}=\sigma _{y}p_{x}$, is not $%
\mathcal{PT}$-symmetric, but the remedy is simple, since the two
representations are unitarily equivalent: the unitary transformation%
\begin{equation}
U=e^{i\frac{\pi }{4}\sigma _{z}}e^{i\frac{\pi }{4}\sigma _{y}}=\frac{1}{2}%
\left[ 1+i\left( \sigma _{x}+\sigma _{y}+\sigma _{z}\right) \right] =\frac{1%
}{2}\left( 
\begin{array}{cc}
1+i & 1+i \\ 
-1+i & 1-i%
\end{array}%
\right)  \label{unitr}
\end{equation}%
changes the $\alpha _{x}$ and $\beta $ matrices of the representation of Ref.%
\cite{SR05} into those of the Dirac representation, since%
\begin{equation*}
\begin{array}{c}
U\sigma _{y}U^{-1}=\sigma _{x}\;, \\ 
U\sigma _{x}U^{-1}=\sigma _{z}\;.%
\end{array}%
\end{equation*}

It is also of interest to determine the operator that corresponds in the
present representation to to the $\mathcal{PT}$ operator (\ref{PT}) of the
Dirac representation: let us notice that $\mathcal{PT}$ \ anticommutes with $%
i\sigma _{x}$, since $\mathcal{T}$ is antilinear, and that $(\mathcal{PT}%
i\sigma _{x})^{2}=1_{2}$, the $2\times 2$ identity matrix. It is not
difficult to check that%
\begin{equation*}
U\mathcal{PT}i\sigma _{x}U^{-1}=-iU\sigma _{x}U^{T}\mathcal{PT=PT\;}.
\end{equation*}

Therefore, $\mathcal{PT}i\sigma _{x}$ is obtained from $\mathcal{PT}$ \ by
means of the similarity transformation that connects matrices in the present
representation with those in the Dirac representation. Moreover, once the
Dirac equation (\ref{scal_eq}) has been solved in the new representation,
and the spinor $\Psi \left( x\right) $ is known, the corresponding solution
in the $\mathcal{PT}$-symmetric Dirac representation will be 
\begin{equation}
\Psi _{D}\left( x\right) =U\Psi \left( x\right) \;.  \label{Psitr}
\end{equation}

As a consequence, if $\Psi _{D}\left( x\right) $ is an eigenstate of $%
\mathcal{PT}$, $\Psi \left( x\right) $ is an eigenstate of $\mathcal{PT}%
i\sigma _{x}$.

In the new representation the two equations satisfied by the components, $%
\psi _{1}\left( x\right) $ and $\psi _{2}\left( x\right) $, of $\Psi \left(
x\right) $%
\begin{equation*}
\left[ 
\begin{array}{cc}
-E & -\frac{d}{dx}+m+S\left( x\right) \\ 
\frac{d}{dx}+m+S\left( x\right) & -E%
\end{array}%
\right] \left( 
\begin{array}{c}
\psi _{1}\left( x\right) \\ 
\psi _{2}\left( x\right)%
\end{array}%
\right) =\left( 
\begin{array}{c}
0 \\ 
0%
\end{array}%
\right) \;,
\end{equation*}%
are easily decoupled as%
\begin{equation}
\left\{ 
\begin{array}{c}
-\frac{1}{2m}\frac{d^{2}}{dx^{2}}\psi _{1}\left( x\right) +\frac{1}{2m}%
\left( \left( m+S\left( x\right) \right) ^{2}-m^{2}-\frac{d}{dx}S\left(
x\right) \right) \psi _{1}\left( x\right) =\frac{E^{2}-m^{2}}{2m}\psi
_{1}\left( x\right) \equiv \epsilon \psi _{1}\left( x\right) \\ 
\psi _{2}\left( x\right) =\frac{1}{E}\left( \frac{d}{dx}+m+S\left( x\right)
\right) \psi _{1}\left( x\right)%
\end{array}%
\right. \;,  \label{scal_sys}
\end{equation}%
where the equation satisfied by $\psi _{1}\left( x\right) $ is Schr\H{o}%
dinger-like, with an effective potential%
\begin{equation}
U_{1}\left( x\right) =\frac{1}{2m}\left( \left( m+S\left( x\right) \right)
^{2}-m^{2}-\frac{d}{dx}S\left( x\right) \right) \;.  \label{U_1_scal}
\end{equation}

If one derives instead the equation satisfied by $\psi _{2}\left( x\right) $%
, the result is a Schr\H{o}dinger-like equation with the same effective
energy $\epsilon =\frac{E^{2}-m^{2}}{2m}$ and an effective potential%
\begin{equation}
U_{2}\left( x\right) =\frac{1}{2m}\left( \left( m+S\left( x\right) \right)
^{2}-m^{2}+\frac{d}{dx}S\left( x\right) \right) \;.  \label{U_2_scal}
\end{equation}

We are thus led again to a pseudo-supersymmetry\cite{SR05}, like in the case
of a pseudoscalar potential. Note that $S\left( x\right) +m\equiv m\left(
x\right) $, the effective Dirac mass, now plays the role of a
superpotential. The supercharges now are%
\begin{equation*}
L=\frac{1}{\sqrt{2m}}\left( -\frac{d}{dx}+m+S\left( x\right) \right) \;,\;M=%
\frac{1}{\sqrt{2m}}\left( \frac{d}{dx}+m+S\left( x\right) \right)
\end{equation*}%
and the partner Hamiltonians%
\begin{equation*}
H_{1}=LM=-\frac{1}{2m}\frac{d^{2}}{dx^{2}}+U_{1}\left( x\right)
\;,\;H_{2}=ML=-\frac{1}{2m}\frac{d^{2}}{dx^{2}}+U_{2}\left( x\right) \;.
\end{equation*}

In this representation, the asymptotic Dirac equation%
\begin{equation*}
\left[ 
\begin{array}{cc}
-E & -\frac{d}{dx}+m+S_{\pm } \\ 
\frac{d}{dx}+m+S_{\pm } & -E%
\end{array}%
\right] \left( 
\begin{array}{c}
\psi _{1\pm }\left( x\right) \\ 
\psi _{2\pm }\left( x\right)%
\end{array}%
\right) =\left( 
\begin{array}{c}
0 \\ 
0%
\end{array}%
\right) \;,
\end{equation*}%
where, as before, $\lim_{x\rightarrow \pm \infty }S\left( x\right) =S_{\pm }$%
, $\lim_{x\rightarrow \pm \infty }\psi _{k}\left( x\right) =\psi _{k\pm
}\left( x\right) $, is easily solved in the form%
\begin{eqnarray*}
\Psi _{\pm }\left( x\right) &=&\left( 
\begin{array}{c}
\psi _{1\pm }\left( x\right) \\ 
\psi _{2\pm }\left( x\right)%
\end{array}%
\right) \\
&=&\widetilde{A}_{\pm }\left( 
\begin{array}{c}
1 \\ 
\widetilde{C}_{\pm }%
\end{array}%
\right) e^{ik_{\pm }x}+\widetilde{B}_{\pm }\left( 
\begin{array}{c}
1 \\ 
\widetilde{D}_{\pm }%
\end{array}%
\right) e^{-ik_{\pm }x}\;,
\end{eqnarray*}%
where $k_{\pm }=\left( E^{2}-\left( m+S_{\pm }\right) ^{2}\right) ^{1/2}$
are the complex asymptotic momenta and 
\begin{equation*}
\begin{array}{c}
\widetilde{C}_{\pm }=\frac{ik_{\pm }+m+S_{\pm }}{E} \\ 
\widetilde{D}_{\pm }=\frac{-ik_{\pm }+m+S_{\pm }}{E}%
\end{array}%
\;,
\end{equation*}%
while $\widetilde{A}_{\pm }$ and $\widetilde{B}_{\pm }$ are to be fixed on
boundary conditions. Owing to the fact that $\Psi _{\pm }\left( x\right) $
and $\Psi _{D\pm }\left( x\right) $ are connected by Eqs. (\ref{unitr}-\ref%
{Psitr}), the asymptotic expansion coefficients $\widetilde{A}_{\pm }$ and $%
\widetilde{B}_{\pm }$ are related by the inverse of formula (\ref{Psitr}), 
\textit{i.e. }$\Psi \left( x\right) =U^{\dag }\Psi _{D}\left( x\right) $, to
the corresponding ones in the Dirac representations, $A_{\pm }$ and $B_{\pm
} $, given in formulae (\ref{psi_trial}) with $V_{\pm }=P_{\pm }=0$, in the
following way%
\begin{equation}
\begin{array}{c}
\widetilde{A}_{\pm }=\frac{A_{\pm }}{\sqrt{2}}e^{-i\frac{\pi }{4}}\frac{%
m+S_{\pm }+E-ik_{\pm }}{m+S_{\pm }+E}\;, \\ 
\widetilde{B}_{\pm }=\frac{B_{\pm }}{\sqrt{2}}e^{-i\frac{\pi }{4}}\frac{%
m+S_{\pm }+E+ik_{\pm }}{m+S_{\pm }+E}\;.%
\end{array}
\label{A_B_tilde}
\end{equation}

It is worthwhile to stress once again that in the present representation of
the Dirac equation $\Psi _{\pm }\left( x\right) $ is not eigenstate of $%
\mathcal{PT}$.

Before discussing specific examples, it is worthwhile to recall that a
reflectionless potential can be obtained as a supersymmetry partner of the
constant potential: in fact, putting $U_{1}\left( x\right) =c$ in formula (%
\ref{U_1_scal}), one obtains a Riccati equation for the superpotential $%
S\left( x\right) +m$, which can be solved by separation of variables, giving
rise to different solutions in connection with the sign of the constant $m+2c
$. The superpotential is a trigonometric function of $x$ when $m+2c<0$ (or $%
c<-\frac{m}{2}$) and a hyperbolic function when $m+2c>0$; consequently, $%
U_{2}\left( x\right) $ is a trigonometric P\"{o}schl-Teller potential in the
former case and a hyperbolic P\"{o}schl-Teller potential in the latter. In
the limiting case $m+2c=0$, one obtains $S\left( x\right) +m=-\frac{1}{x+d}$%
, where $d$ is an arbitrary constant, and $U_{2}\left( x\right) =\frac{1}{m}%
\frac{1}{\left( x+d\right) ^{2}}-\frac{m}{2}$. When $d=i\epsilon $, with
real $\epsilon $, $U_{2}\left( x\right) $ is qualitatively similar, apart
from the additive constant, to the potential studied in Section 3. Similar
considerations could be made in the case of the supersymmetry involving
pseudoscalar potentials presented in Section 4.

A simple example of reflectionless scalar potential is the $\mathcal{PT}$%
-symmetrized form of the real potential with one bound state derived in Ref.%
\cite{TNZ93}%
\begin{eqnarray}
S\left( x\right) &=&-\frac{2\kappa _{B}^{2}}{m+E_{B}\cosh \left( 2\kappa
_{B}\left( x+i\epsilon \right) \right) }  \label{S_1} \\
&=&-\frac{\kappa _{B}^{2}}{E_{B}}\frac{1}{\cosh \left( \kappa _{B}\left(
x+i\epsilon \right) -\lambda _{B}\right) \cosh \left( \kappa _{B}\left(
x+i\epsilon \right) +\lambda _{B}\right) }\;.  \notag
\end{eqnarray}

Here, $E_{B}=\frac{2m}{\sqrt{c_{S}^{2}+4}}$ and $\kappa _{B}=\sqrt{%
m^{2}-E_{B}^{2}}=\frac{c_{S}m}{\sqrt{c_{S}^{2}+4}}$ are energy and momentum
of the bound state, expressed as functions of the coupling strength $c_{S}$
in the auxiliary non-linear Dirac equation discussed in the above mentioned
reference \ and $\lambda _{B}=\frac{1}{2}$arccosh$\left( \frac{m}{E_{B}}%
\right) $. The two partner potentials from formulae (\ref{U_1_scal}-\ref%
{U_2_scal}) can be written in the compact form%
\begin{equation}
\begin{array}{c}
U_{1}\left( x\right) =-\frac{\kappa _{B}^{2}}{m\cosh ^{2}\left( \kappa
_{B}\left( x+i\epsilon \right) -\lambda _{B}\right) }\;, \\ 
U_{2}\left( x\right) =-\frac{\kappa _{B}^{2}}{m\cosh ^{2}\left( \kappa
_{B}\left( x+i\epsilon \right) +\lambda _{B}\right) }\;.%
\end{array}
\label{U_1_2_scal}
\end{equation}

\ While $S\left( x\right) $ is $\mathcal{PT}$-symmetric, the partner
potentials (\ref{U_1_2_scal}) are not, as expected from Eqs. (\ref{U_1_scal}-%
\ref{U_2_scal}). The two Schr\H{o}dinger-like equations with potentials (\ref%
{U_1_2_scal})%
\begin{equation}
-\frac{1}{2m}\frac{d^{2}}{dx^{2}}\psi _{i}\left( x\right) +U_{i}\left(
x\right) \psi _{i}\left( x\right) =\varepsilon \psi _{i}\left( x\right)
\;\;\left( i=1,2\right)  \label{Schr_eq_1_2}
\end{equation}%
are satisfied by the bound-state wave functions%
\begin{equation}
\begin{array}{c}
\psi _{1}\left( x\right) =\frac{N_{1}}{\cosh \left( \kappa _{B}\left(
x+i\epsilon \right) -\lambda _{B}\right) } \\ 
\psi _{2}\left( x\right) =\frac{N_{2}}{\cosh \left( \kappa _{B}\left(
x+i\epsilon \right) +\lambda _{B}\right) }%
\end{array}
\label{psi_bs_1_2}
\end{equation}%
respectively, where $N_{1}$ and $N_{2}$ are normalization constants. Note
that the two partners have the same discrete spectrum, \textit{i.e.} one
bound state with real energy $\varepsilon =-\frac{\kappa _{B}^{2}}{2m}$: the
pseudosupersymmetry is thus spontaneously broken\cite{NT93}. The two
normalization constants are related by the second Eq. (\ref{scal_sys}):
after replacing in it formula (\ref{S_1}) for $S\left( x\right) $ and the
first formula (\ref{psi_bs_1_2}) for $\psi _{1}\left( x\right) $, some
straightforward algebra gives $N_{2}=N_{1}$. The corresponding Dirac spinor, 
$\Psi \left( x\right) =\left( 
\begin{array}{c}
\psi _{1}\left( x\right) \\ 
\psi _{2}\left( x\right)%
\end{array}%
\right) $, has real energy $E_{B}=\sqrt{m^{2}-\kappa _{B}^{2}}$ and is
normalized to one. Hence, 
\begin{eqnarray*}
1 &=&\int\limits_{-\infty }^{+\infty }\Psi ^{\dag }\left( x\right) \Psi
\left( x\right) dx=\int\limits_{-\infty }^{+\infty }\left\vert \psi
_{1}\left( x\right) \right\vert ^{2}dx+\int\limits_{-\infty }^{+\infty
}\left\vert \psi _{2}\left( x\right) \right\vert ^{2}dx \\
&=&\frac{4}{\kappa _{B}}\left\vert N_{1}\right\vert ^{2}\int\limits_{-\infty
}^{+\infty }\frac{dy}{\cosh \left( 2y\right) +\cos \left( 2\kappa
_{B}\epsilon \right) }=\frac{4\epsilon \left\vert N_{1}\right\vert ^{2}}{%
\sin \left( \kappa _{B}\epsilon \right) \cos \left( \kappa _{B}\epsilon
\right) } \\
&\Rightarrow& \left\vert N_{1}\right\vert ^{2}=\frac{\sin \left( \kappa
_{B}\epsilon \right) \cos \left( \kappa _{B}\epsilon \right) }{4\epsilon }\;.
\end{eqnarray*}

Eqs. (\ref{Schr_eq_1_2}) with $\varepsilon >0$ are satisfied by the
scattering wave functions%
\begin{equation}
\begin{array}{c}
\psi _{1}\left( x\right) =\frac{e^{ik(x+i\epsilon )}}{ik+\kappa _{b}}\left[
ik-\kappa _{b}\tanh \left( \kappa _{b}\left( x+i\epsilon \right) -\lambda
_{B}\right) \right] \\ 
\psi _{2}\left( x\right) =\frac{e^{ik(x+i\epsilon )}}{ik+\kappa _{b}}\left[
ik-\kappa _{b}\tanh \left( \kappa _{b}\left( x+i\epsilon \right) +\lambda
_{B}\right) \right]%
\end{array}
\label{psi_scatt_1_2}
\end{equation}%
respectively, with $k=\sqrt{2m\varepsilon }$, if boundary conditions for
incident progressive waves ($L\rightarrow R$) are imposed. Since $%
\lim_{x\rightarrow \pm \infty }\tanh \left( \kappa _{b}\left( x+i\epsilon
\right) \pm \lambda _{B}\right) =\pm 1$, $\psi _{i}\left( x\right) $ does
not contain a reflected component, so that $R_{L\rightarrow R}^{\left(
i\right) }=0$. It is also immediate to determine the transmission
coefficient 
\begin{equation}
T_{L\rightarrow R}^{\left( i\right) }=\frac{\lim_{x\rightarrow +\infty }\psi
_{i}\left( x\right) }{\lim_{x\rightarrow -\infty }\psi _{i}\left( x\right) }=%
\frac{ik-\kappa _{b}}{ik+\kappa _{b}}\;.  \label{T_LR_1_2}
\end{equation}

Since $T_{L\rightarrow R}^{\left( 1\right) }=T_{L\rightarrow R}^{\left(
2\right) }$, formula (\ref{T_LR_1_2}) yields the transmission coefficient $%
T_{L\rightarrow R}$ of the Dirac spinor, which has unit modulus, as
expected. The replacement $k\rightarrow -k$ changes progressive waves (\ref%
{psi_scatt_1_2}) into regressive waves, whose transmission coefficient now is%
\begin{equation}
T_{R\rightarrow L}^{\left( i\right) }=\frac{\lim_{x\rightarrow -\infty }\psi
_{i}\left( x\right) }{\lim_{x\rightarrow +\infty }\psi _{i}\left( x\right) }=%
\frac{-ik+\kappa _{b}}{-ik-\kappa _{b}}=T_{L\rightarrow R}^{\left( i\right) }
\label{T_RL_1_2}
\end{equation}%
with $R_{R\rightarrow L}^{\left( i\right) }=R_{L\rightarrow R}^{\left(
i\right) }=0$. The equality of the transmission coefficients could have been
proved also by means of formula (\ref{Ratio}), since, in the present case, $%
C_{-}=C_{+}=\frac{ik+m}{E}$ and $D_{-}=D_{+}=\frac{-ik+m}{E}$.

\section{Comments and outlook}

In the present work we have extended the analysis of reflectionless $%
\mathcal{PT}$-symmetric potentials in non-relativistic quantum mechanics
presented in Ref.\cite{CDV07} to relativistic quantum mechanics with
different forms of potentials in the Dirac equation: scalar, pseudoscalar ,
or a mixture of scalar and vector potentials. We have examined the
connection between reflectionlessness and exact $\mathcal{PT}$ symmetry of
bound-state wave functions and asymptotic wave functions even in the case
the potentials have non-zero limits at $x\rightarrow \pm \infty $, thus
removing a constraint imposed in Ref.\cite{CDV07}. Along this line, one
could study further reflectionless potentials that diverge at $x\rightarrow
\pm \infty $ , such as the class of real symmetric potentials $V\left(
x\right) =-x^{2k+2}$ ($k=1,2,...$) discussed in Ref.\cite{ABB05}. A main
drawback is that they are not exactly solvable in general and one has to
resort to some approximate method, such as the WKB method of Ref.\cite{ABB05}%
.

Reflectionless potentials are expected to play a peculiar role also in $%
\mathcal{PT}$-symmetric quantum mechanics in higher dimensions, very little
explored up to the present time and only in non-relativistic problems. In
three dimensions, for instance, only non-central potentials can exhibit
non-trivial $\mathcal{PT}$ symmetry: in polar coordinates $r$, $\theta $, $%
\phi $, they satisfy the relation\cite{Le07}%
\begin{equation*}
V\left( \mathbf{r}\right) \equiv V\left( r,\theta ,\phi \right) =V^{\ast
}\left( r,\pi -\theta ,\phi +\pi \right) =V^{\ast }\left( -\mathbf{r}\right)
\;.
\end{equation*}

General characteristics of bound states are discussed in Ref.\cite{Le07} in
case of exact $\mathcal{PT}$ symmetry and in Ref.\cite{Le08} when the
symmetry is spontaneously broken.

Scattering states of $\mathcal{PT}$-symmetric potentials in higher
dimensions have not been discussed so far, but a complete analytic
description should be possible if these non-central transparent potentials
can be related to the euclidean group in $n$ dimensions, $E\left( n\right) $%
, for $n>3$, since it is the maximal symmetry group of the transparent null
potential in $n$ dimensions. For instance, Ref.\cite{Ke07} solves two
classes of real transparent potentials in three dimensions that admit $%
E\left( 4\right) $ as the potential group, in the sense that the
corresponding Hamiltonians depend on the restriction of the quadratic
Casimir operator to subspaces appearing in two subgroup reduction chains of $%
E\left( 4\right) $. \ The possible group-theoretical treatment of complex
potentials would require non-standard realizations of the non-compact Lie
group involved (non-unitary representations), along the lines of Ref.\cite%
{Le99}. Extension of this study to relativistic quantum mechanics would be
possible by relating to the Casimir operator of a non-compact group the mass
invariant operator of the system within the framework of a Bakamjian-Thomas
realization of the generators of the Poincar\'{e} group, under which one has
to ensure the invariance of the scattering matrix\cite{KV08}.

Finally, it is worthwhile to mention that interest in the first quantized
version of the Dirac equation has been renewed in recent years by quantum
simulations of the dynamics of Dirac fermions with controllable laboratory
systems underlying the same mathematical models: single trapped ions are
particularly suited to this purpose and experiments have been designed to
simulate the one-dimensional trembling motion (Zitterbewegung\cite{Gr97},%
\cite{CF91}) of a free Dirac fermion\cite{La07},\cite{Ge09}, the
overcriticality effects of a vector potential well in one dimension\cite%
{La07}, the relativistic Landau levels generated in a constant homogeneous
magnetic field\cite{Be07a}, the pseudoscalar Dirac oscillator in two
dimensions\cite{Be07b}. The possibility of simulating in quantum optical
systems the one-dimensional Dirac dynamics in presence of non-Hermitian
potentials, such as the $\mathcal{PT}$-symmetric ones considered in the
present work, would be of mutual benefit to both fields of research.

\begin{acknowledgement}
We are grateful to an anonymous referee for pointing out the interest of the
low-dimensional Dirac equation within the novel field of quantum simulations.
We are also grateful to M. S. Plyushchay for informing us that the 
Bogoliubov-de Gennes Hamiltonian\cite{Co09} can lead to a Dirac-like 
effective equation
with underlying supersymmetry similar to those of Sections 4 and 5 of the 
present work.
\end{acknowledgement}

\appendix
\section{Relations between transmission and reflection coefficients of
supersymmetry partners in the pseudoscalar case}

Let us consider a scattering solution of one of the two partner
Hamiltonians, \textit{e. g. }$H_{1}$, with energy $\epsilon \equiv \frac{%
E^{2}-m^{2}}{2m}\geq 0$%
\begin{equation*}
H_{1}\phi ^{\left( 1\right) }\left( x\right) =\epsilon \phi ^{\left(
1\right) }\left( x\right)
\end{equation*}%
and a solution, $\phi ^{\left( 2\right) }$, \ of Hamiltonian $H_{2}$ with
the same energy $\epsilon $%
\begin{equation*}
H_{2}\phi ^{\left( 2\right) }\left( x\right) =\epsilon \phi ^{\left(
2\right) }\left( x\right) \;.
\end{equation*}

Since $H_{1}=LM$ and $H_{2}=ML$, where $L$ and $M$ are the differential
operators defined in formulae (\ref{L_M}), we immediately see that $\epsilon
L\phi ^{\left( 2\right) }=LH_{2}\phi ^{\left( 2\right) }=LML\phi ^{\left(
2\right) }=H_{1}L\phi ^{\left( 2\right) }$, which means that $L\phi ^{\left(
2\right) }$ must be proportional to $\phi ^{\left( 1\right) }$, or 
\begin{equation*}
\phi ^{\left( 1\right) }\left( x\right) =\mathcal{C}L\phi ^{\left( 2\right)
}\left( x\right) \;,
\end{equation*}%
where $\mathcal{C}$ is a constant to be determined. Let us assume now that $%
\phi ^{\left( 2\right) }\left( x\right) $ is a progressive wave, with
asymptotic behaviour%
\begin{equation*}
\begin{array}{c}
\lim_{x\rightarrow -\infty }\phi ^{\left( 2\right) }\left( x\right)
=e^{ik_{-}x}+R_{L\rightarrow R}^{\left( 2\right) }e^{-ik_{-}x}\;, \\ 
\lim_{x\rightarrow +\infty }\phi ^{\left( 2\right) }\left( x\right)
=T_{L\rightarrow R}^{\left( 2\right) }e^{ik_{+}x}\;,%
\end{array}%
\end{equation*}%
where $k_{\pm }=\sqrt{2m\left( \epsilon -U_{2}\left( \pm \infty \right)
\right) }=\sqrt{E^{2}-m^{2}-P_{\pm }^{2}}$ are the asymptotic momenta. Thus,
we have%
\begin{equation*}
\begin{array}{c}
\lim_{x\rightarrow -\infty }\phi ^{\left( 1\right) }\left( x\right) =%
\mathcal{C}\lim_{x\rightarrow -\infty }L\phi ^{\left( 2\right) }\left(
x\right) =\frac{\mathcal{C}}{\sqrt{2m}}\left( \frac{d}{dx}+P_{-}\right)
\left( e^{ik_{-}x}+R_{L\rightarrow R}^{\left( 2\right) }e^{-ik_{-}x}\right)
\\ 
=\frac{\mathcal{C}}{\sqrt{2m}}\left[ \left( ik_{-}+P_{-}\right)
e^{ik_{-}x}+\left( -ik_{-}+P_{-}\right) R_{L\rightarrow R}^{\left( 2\right)
}e^{-ik_{-}x}\right]%
\end{array}%
\end{equation*}%
and 
\begin{equation*}
\begin{array}{c}
\lim_{x\rightarrow +\infty }\phi ^{\left( 1\right) }\left( x\right) =%
\mathcal{C}\lim_{x\rightarrow +\infty }L\phi ^{\left( 2\right) }\left(
x\right) =\frac{\mathcal{C}}{\sqrt{2m}}\left( \frac{d}{dx}+P_{+}\right)
T_{L\rightarrow R}^{\left( 2\right) }e^{ik_{+}x} \\ 
=\frac{\mathcal{C}}{\sqrt{2m}}\left( ik_{+}+P_{+}\right) T_{L\rightarrow
R}^{\left( 2\right) }e^{ik_{+}x}\;.%
\end{array}%
\end{equation*}

On the other hand, we know that%
\begin{equation*}
\begin{array}{c}
\lim_{x\rightarrow -\infty }\phi ^{\left( 1\right) }\left( x\right)
=e^{ik_{-}x}+R_{L\rightarrow R}^{\left( 1\right) }e^{-ik_{-}x}\;, \\ 
\lim_{x\rightarrow +\infty }\phi ^{\left( 1\right) }\left( x\right)
=T_{L\rightarrow R}^{\left( 1\right) }e^{ik_{+}x}\;.%
\end{array}%
\end{equation*}

Hence we obtain formulae (\ref{T12_R_12}) of the text%
\begin{equation*}
\begin{array}{c}
\frac{\mathcal{C}}{\sqrt{2m}}=\frac{1}{ik_{-}+P_{-}} \\ 
R_{L\rightarrow R}^{\left( 1\right) }=\frac{-ik_{-}+P_{-}}{ik_{-}+P_{-}}%
R_{L\rightarrow R}^{\left( 2\right) } \\ 
T_{L\rightarrow R}^{\left( 1\right) }=\frac{ik_{+}+P_{+}}{ik_{-}+P_{-}}%
T_{L\rightarrow R}^{\left( 2\right) }%
\end{array}%
\end{equation*}

One proceeds in a similar way for regressive waves, $\phi ^{\left( 1\right)
}\left( x\right) =\mathcal{C}^{\prime }L\phi ^{\left( 2\right) }\left(
x\right) $, with asymptotic behaviour%
\begin{equation*}
\begin{array}{c}
\lim_{x\rightarrow -\infty }\phi ^{\left( j\right) }\left( x\right)
=T_{R\rightarrow L}^{\left( j\right) }e^{-ik_{-}x}\;, \\ 
\lim_{x\rightarrow +\infty }\phi ^{\left( j\right) }\left( x\right)
=e^{-ik_{+}x}+R_{R\rightarrow L}^{\left( j\right) }e^{ik_{+}x}\;,%
\end{array}%
\;(j=1,2)
\end{equation*}%
with the result%
\begin{equation*}
\begin{array}{c}
\frac{\mathcal{C}^{\prime }}{\sqrt{2m}}=\frac{1}{-ik_{+}+P_{+}} \\ 
R_{R\rightarrow L}^{\left( 1\right) }=\frac{ik_{+}+P_{+}}{-ik_{+}+P_{+}}%
R_{R\rightarrow L}^{\left( 2\right) } \\ 
T_{R\rightarrow L}^{\left( 1\right) }=\frac{-ik_{-}+P_{-}}{-ik_{+}+P_{+}}%
T_{R\rightarrow L}^{\left( 2\right) }%
\end{array}%
\;,
\end{equation*}%
corrresponding to formulae (\ref{T12_R_12_b}) of the text.

\end{document}